\documentclass[10pt]{article}
\RequirePackage{lhchiggs}
\RequirePackage{cernunits}
\RequirePackage{heppennames2}
\usepackage{texnames}
\usepackage[T1]{fontenc}
\usepackage{color}
\usepackage{epsfig}
\usepackage{axodraw}
\usepackage{epsfig}
\usepackage{graphicx}
\usepackage{rotate}
\usepackage{latexsym}
\usepackage{amssymb}
\graphicspath{{./}{figures/}}
\newcommand{\FR}{\rm{\scriptscriptstyle{FR}}}
\newcommand{\myNLO}{\rm{\scriptscriptstyle{NLO}}}
\newcommand{\mySM}{\rm{\scriptscriptstyle{SM}}}
\newcommand{\myEW}{{\scriptscriptstyle{E}\scriptscriptstyle{W}}}

\newcommand{\ssQ}{{\scriptscriptstyle{Q}}}
\newcommand{\ssF}{{\scriptscriptstyle{F}}}

\newcommand{\ssL}{{\scriptscriptstyle{L}}}

\newcommand{\bqas}{\begin{eqnarray*}}
\newcommand{\eqas}{\end{eqnarray*}}
\newcommand{\nl}{\nonumber\\}

\newcommand{\lpar}{\left(}                            % bracketing
\newcommand{\rpar}{\right)}

\newcommand{\bq}{\begin{equation}}                    % equationing
\newcommand{\eq}{\end{equation}}
\newcommand{\bqa}{\arraycolsep 0.14em\begin{eqnarray}}
\newcommand{\eqa}{\end{eqnarray}}
\newcommand{\ba}[1]{\begin{array}{#1}}
\newcommand{\ea}{\end{array}}
\newcommand{\ben}{\begin{enumerate}}
\newcommand{\een}{\end{enumerate}}
\newcommand{\bei}{\begin{itemize}}
\newcommand{\eei}{\end{itemize}}
\newcommand{\eqn}[1]{Eq.(\ref{#1})}

\newcommand{\fig}[1]{Fig.~\ref{#1}}

\newcommand{\Mbp}{\mathswitch {m_{\PQbpr}}}
\newcommand{\Mtp}{\mathswitch {m_{\PQtpr}}}
\newcommand{\Mlp}{\mathswitch {m_{\Pl'}}}
\newcommand{\Mnp}{\mathswitch {m_{\PGnl'}}}
\newcommand{\MGq}{\mathswitch {m_{\PQq}}}
\newcommand{\MGqs}{\mathswitch {m^2_{\PQq}}}
\newcommand{\MGl}{\mathswitch {m_{\Pl}}}
\newcommand{\MGf}{\mathswitch {m_{f}}}
\newcommand{\MGfs}{\mathswitch {m^2_{f}}}
\newcommand{\Bref}[1]{Ref.~\cite{#1}}
\newcommand{\Brefs}[1]{Refs.~\cite{#1}}

\newcommand{\twol}{{\mbox{\scriptsize 2-loop}}}

\newcommand{\eg}{e.g.\xspace}
\newcommand{\ie}{i.e.\xspace}

\newcommand{\SPp}{~}       %% space before punctuation in math
\newcommand\pubnumber{% 
MPP-2011-98
                     }
\newcommand\pubdate{August 9, 2011}
\def\csumb{Dipartimento di Fisica Teorica, Universit\`a di Torino, Italy\\
           INFN, Sezione di Torino, Italy}
\def\csumc{Max-Planck-Institut f{\"u}r Physik, 
           (Werner-Heisenberg-Institut), \\
           F{\"o}hringer Ring 6, 
           80805 M{\"u}nchen, 
           Germany}
\def\csumd{
Lehrstuhl f\"ur Theoretische Physik II, 
Institut f\"ur Theoretische Physik und Astrophysik,\\
Universit\"at W\"urzburg,
97074 W\"urzburg, Germany}
\def\support{\footnote{Work supported by MIUR under contract
2008H8F9RA$\_$002.}}
\def\Title#1{\begin{center} {\Large\bf #1 } \end{center}}

\def\Author#1{\begin{center}{ \sc #1} \end{center}}
\def\Address#1{\begin{center}{ \it #1} \end{center}}

\newcommand\pubblock{\rightline{\begin{tabular}{l} \pubnumber\\
         \pubdate\\  \end{tabular}}}
\newenvironment{Abstract}{\begin{quotation}  }{\end{quotation}}

\def\Acknowledgments{\bigskip  \bigskip \begin{center}
          \large\bf Acknowledgments\end{center}}
\def\email#1{\footnote{#1}}
\makeatletter
\def\section{\@startsection{section}{0}{\z@}{5.5ex plus .5ex minus
 1.5ex}{2.3ex plus .2ex}{\large\bf}}
\def\subsection{\@startsection{subsection}{1}{\z@}{3.5ex plus .5ex minus
 1.5ex}{1.3ex plus .2ex}{\normalsize\bf}}
\def\subsubsection{\@startsection{subsubsection}{2}{\z@}{-3.5ex plus
-1ex minus  -.2ex}{2.3ex plus .2ex}{\normalsize\sl}}

\renewcommand{\@makecaption}[2]{%
   \vskip 10pt
   \setbox\@tempboxa\hbox{\small #1: #2}
   \ifdim \wd\@tempboxa >\hsize     % IF longer than one line:
       \small #1: #2\par          %   THEN set as ordinary paragraph.
     \else                        %   ELSE  center.
       \hbox to\hsize{\hfil\box\@tempboxa\hfil}
   \fi}
 \def\citenum#1{{\def\@cite##1##2{##1}\cite{#1}}}
\def\citea#1{\@cite{#1}{}}
\newcount\@tempcntc
\def\@citex[#1]#2{\if@filesw\immediate\write\@auxout{\string\citation{#2}}\fi
  \@tempcnta\z@\@tempcntb\m@ne\def\@citea{}\@cite{\@for\@citeb:=#2\do
    {\@ifundefined
       {b@\@citeb}{\@citeo\@tempcntb\m@ne\@citea\def\@citea{,}{\bf }\@warning
       {Citation `\@citeb' on page \thepage \space undefined}}%
    {\setbox\z@\hbox{\global\@tempcntc0\csname b@\@citeb\endcsname\relax}%
     \ifnum\@tempcntc=\z@ \@citeo\@tempcntb\m@ne
       \@citea\def\@citea{,}\hbox{\csname b@\@citeb\endcsname}%
     \else
      \advance\@tempcntb\@ne
      \ifnum\@tempcntb=\@tempcntc
      \else\advance\@tempcntb\m@ne\@citeo
      \@tempcnta\@tempcntc\@tempcntb\@tempcntc\fi\fi}}\@citeo}{#1}}
\def\@citeo{\ifnum\@tempcnta>\@tempcntb\else\@citea\def\@citea{,}%
  \ifnum\@tempcnta=\@tempcntb\the\@tempcnta\else
  {\advance\@tempcnta\@ne\ifnum\@tempcnta=\@tempcntb \else\def\@citea{--}\fi
    \advance\@tempcnta\m@ne\the\@tempcnta\@citea\the\@tempcntb}\fi\fi}
\makeatother
\begin{document}
\begin{titlepage}
\pubblock
%\pubdate
%
\vfill
\def\thefootnote{\fnsymbol{footnote}}
\Title{Complete Electroweak Corrections to Higgs production\\[0.3cm]
in a Standard Model with four generations at the LHC.
  \footnote[9]{Work performed within the Higgs Cross-Section Working Group\\
{\tt https://twiki.cern.ch/twiki/bin/view/LHCPhysics/CrossSections}.}\support}
\vfill
\Author{Giampiero Passarino 
\email{giampiero@to.infn.it}}               
\Address{\csumb}
\Author{Christian Sturm     
\email{sturm@mpp.mpg.de}}
\Address{\csumc}
\Author{Sandro Uccirati     
\email{Sandro.Uccirati@physik.uni-wuerzburg.de}} 
\Address{\csumd}
\vfill
\vfill
\begin{Abstract}
\noindent 
Complete electroweak two-loop corrections to the process $\Pg \Pg \to
\PH$ are presented and discussed in a Standard Model with a fourth
generation of heavy fermions. The latter is studied at the LHC to put
exclusion limits on a fourth generation of heavy fermions. Therefore
also a precise knowledge of the electroweak(EW)
next-to-leading-order(NLO) corrections is important. The corrections due
to the fourth generation are positive and large for a light Higgs boson,
positive but relatively small around the $\PAQt{-}\PQt$ threshold and
start to become negative for a Higgs boson mass around $\MH = 450\UGeV$.
Increasing further the value of the Higgs boson mass, the EW NLO effects
tend to become huge and negative, ${\cal O}( -100\%)$, around the
heavy-fermion threshold, assumed at $1.2\UTeV$, so that
$\Pg\Pg\,$-fusion becomes non-perturbative. Above that threshold they
start to grow again and become positive around $\MH= 1.75\UTeV$. The
behaviour at even larger values of $\MH$ shows a positive enhancement,
${\cal O}(+100\%)$ at $\MH= 3\UTeV$.
\end{Abstract}
\vfill
\begin{center}
\end{center}
\end{titlepage}
\def\thefootnote{\arabic{footnote}}
\setcounter{footnote}{0}
%--
\small
\thispagestyle{empty}
%\tableofcontents
%
\normalsize
%--
\clearpage
\setcounter{page}{1}
%--
\section{Introduction \label{sec:intro}}
%--
In the last year there have been intensive studies at the LHC aimed to
put exclusion limits on a fourth generation of heavy fermions in the
Standard Model~\cite{Aad:2009wy,Aad:2011qi,Chatrchyan:2011em}.  For
similar searches at Tevatron we refer to \Bref{Aaltonen:2010sv}.
Recently, the overall combination of six Standard Model (SM) Higgs boson
searches has been presented by the CMS Collaboration~\cite{CMSexcl}
using the following Higgs boson decay signatures: $\PH \to \PGg\PGg,
\PGt\PGt, \PW\PW \to 2\Pl\,2\PGn, \PZ\PZ \to 2\Pl\,2\PGn, \PZ \PZ \to
2\Pl\,2\PQq$. The same experimental search results, reinterpreted in the
context of the Standard Model with four fermion generations (SM4), allow
them to exclude the SM4 Higgs boson with a mass in the range
$120{-}600\UGeV$ at $95\%$ C.L. 

From direct searches the situation is as follows: for the decay of a
fourth generation quark $\PQbpr \to \PQt \PW$, the LHC can put a limit
close to $490\,\UGeV$ for $1\,\Ufb^{-1}$ data.  For $\PQbpr \to \PQb
\PW$ and for $\approx 200\,\Upb^{-1}$ they can already obtain a limit
around $420\,\UGeV$ without a full optimization.  The relevant
production channel for these searches is gluon-gluon fusion
($\Pg\Pg$-fusion) that is sensitive to new coloured, heavy
particles. There is little doubt that a Standard Model with a fourth
generation of heavy fermions cannot stimulate great interest (for an
additional vector-like generation see \Bref{Ishiwata:2011hr}), however,
the spectacular modification in the Higgs boson cross-section at hadron
colliders can be tested easily with LHC data.

So far, the experimental analysis has concentrated on models with 
ultra-heavy fourth generation fermions, excluding the possibility that
the Higgs boson decays to heavy neutrinos. Furthermore, the full
two-loop electroweak corrections have been included under the assumption
that they are dominated by light fermions. At the moment the
experimental strategy consists in computing the cross-section ratio $R =
\sigma({\rm SM4})/\sigma({\rm SM3})$ with {\tt
  HIGLU}~\cite{Spira:1996if} while NLO electroweak radiative corrections are
switched off, where ${\rm SM3}$ stands for the Standard Model with three generations.

In this work we concentrate on the computation of the complete two-loop electroweak
corrections and refer to the work
of~\Bref{Anastasiou:2011qw,Anastasiou:2010bt} for the inclusion of the QCD  
corrections. In the next Section~\ref{sec:HeavyFermions} we discuss the limit of heavy
fermions in the gluon fusion channel. In Section~\ref{sec:Results} we present our
results for the electroweak corrections in the ${\rm SM4}$ and discuss in
Section~\ref{sec:Exclusion} the impact of the fourth generations on the
Higgs boson decay. Finally we close with our conclusions in Section~\ref{sec:Conclusion}. 

%--
\section{Heavy fermion mass limit in gluon-gluon fusion\label{sec:HeavyFermions}}
%--
Recently, there have been some confusing and inaccurate statements on
the impact of two-loop electroweak (EW) corrections to Higgs boson
production at the LHC (through $\Pg\Pg$-fusion) in a Standard Model with
a fourth generation of fermions.  The naive expectation is that light
fermions dominate the low Higgs boson mass regime and, therefore,
electroweak corrections can be well approximated by using the available
ones~\cite{Actis:2008ts,Actis:2008ug} in the Standard Model with three
generations (SM3).  It is worth noting that the leading behaviour of the
EW corrections for high values of masses in the fourth generation has
been known for a long time~\cite{Djouadi:1994ge,Djouadi:1997rj} (see
also \Bref{Fugel:2004ug}) showing an enhancement of the radiative
corrections.

To avoid misunderstandings we define the following terminology: for a given 
amplitude $A$, in the limit of the fermion mass $\MGf \to \infty$ we will 
distinguish {\em decoupling} for $A \sim 1/\MGfs$ (or more), {\em screening} 
for $A \to\,$ constant (or $\ln \MGfs$) and {\em enhancement} for $A \sim \MGfs$ 
(or more). 
To discuss decoupling we need few definitions: SM3 is the usual SM with one 
$\PQt{-}\PQb$ doublet; SM4 is the extension of SM3 with a new family of heavy 
fermions with $\PQtpr{-}\PQbpr$ quarks and $\Pl'{-}\PGnl'$ leptons. 
All relevant formulae for the asymptotic limit can be found in 
\Bref{Djouadi:1994ge,Djouadi:1997rj}.
The amplitude for $\Pg\Pg$-fusion is (only EW corrections are considered here)
%--
\bqa
A(\hbox{SM3}) &=& A^{\onel}_{\PQt} + A^{\myNLO}_3, \quad
A^{\myNLO}_3 = A^{\twol}_{\PQt} + \delta^{\FR}_{\PQt}\,A^{\onel}_{\PQt},
\nl
A(\hbox{SM4}) &=& A^{\onel}_{\ssQ} + A^{\myNLO}_4, \quad
A^{\myNLO}_4 = A^{\twol}_{\ssQ} + \delta^{\FR}_{\ssQ + \ssL}\,A^{\onel}_{\ssQ},
\label{defA}
\eqa
%--
where 
%--
\bqa
A_{\ssQ} = A_{\PQt + \PQtpr + \PQbpr}, \qquad 
\delta^{\FR}_{\ssQ + \ssL} = \delta^{\FR}_{\PQt + \PQtpr + \PQbpr + \Pl' + \PGnl'}\SPp.
\eqa
%--
In \eqn{defA} $\delta^{\FR}$ gives the contribution from finite renormalization, 
including Higgs boson wave-function renormalization (see Sect.~$3.4$ of
\Bref{Actis:2008ts} for technical details).

First we recall the standard argument for asymptotic behaviour in the lowest-order (LO) 
$\Pg\Pg$-fusion process, extendible to next-to-leading-order (NLO) and
next-to-next-to-leading-order (NNLO) QCD corrections~\cite{Spira:1995rr}
and give a simple argument to prove enhancement at the NLO EW level.

Any Feynman diagram contributing to the Higgs-gluon-gluon vertex has dimension 
one; however, the total Higgs-gluon-gluon amplitude must be proportional to
$T_{\mu\nu} = p^2\,\delta_{\mu\nu} - p_{\mu} p_{\nu}$ (where $p$ is the Higgs 
momentum) because of gauge invariance. 
For any fermion $f$ the Yukawa coupling is proportional to $\MGf/\MW$ and $T$ has 
dimension two; therefore, the asymptotic behaviour of any diagram must be 
proportional to $T/\MGf$ when $\MGf \to \infty$.
The part of the diagram which is not proportional to $T$ will cancel in the total
because of gauge invariance (all higher powers of $\MGf$ will go away and this 
explains the presence of huge cancellations in the total amplitude).
At LO there is only one Yukawa coupling as in NLO(NNLO) QCD where one adds only 
gluon lines, so there is screening.
At the EW NLO there are diagrams with three Yukawa couplings, therefore giving 
the net $\MGfs$ behaviour predicted in~\cite{Djouadi:1994ge}, so there is 
enhancement and at two-loop level it goes at most with $\MGfs$. 
To be more precise, let us define
%--
\bq
\sigma_{\mySM 3}\lpar \Pg\Pg \to \PH\rpar = 
\sigma^{\LO}_{\mySM 3}\lpar \Pg\Pg \to \PH\rpar\,\Bigl( 1 + \delta^3_{\myEW}\Bigr)
\quad\mbox{and}\quad
\sigma_{\mySM 4}\lpar \Pg\Pg \to \PH\rpar = 
\sigma^{\LO}_{\mySM 4}\lpar \Pg\Pg \to \PH\rpar\,\Bigl( 1 + \delta^4_{\myEW}\Bigr).
\eq
%--
Analysing the results of~\Bref{Djouadi:1994ge,Djouadi:1997rj}, which is valid 
for a light Higgs boson, one can see that in SM3 there is enhancement in the 
quark sector for $\Mt \gg \Mb$ ($\delta^3_{\myEW} \sim G_{\ssF}\,m^2_{\PQt}$ where 
$G_{\ssF}$ is the Fermi coupling constant).
The full calculation of~\Bref{Actis:2008ug} shows that the physical value for the 
top-quark mass is not large enough to make this quadratic behaviour relevant with 
respect to the contribution from light fermions in a wide range of the
Higgs boson mass.
From~\Bref{Djouadi:1994ge} one can also understand that an hypothetical
SM3 with mass-degenerate $\PQt{-}\PQb$ quarks ($\Mt = \Mb = \MGq$) would generate 
an enhancement in the small mass region with the opposite sign 
($\delta^3_{\myEW} \sim -\,G_{\ssF}\,\MGqs$).
Moving to SM4, Eq.(62) of~\Bref{Djouadi:1997rj} shows that the enhanced terms 
coming from finite renormalization exactly cancel the similar contribution from 
two-loop diagrams for $\Mtp= \Mbp$, so that for mass-degenerate quarks 
$\PQtpr{-}\PQbpr$ we observe screening in the fourth generation quark sector.
This accidental cancellation follows from the $3$ of colour $SU(3)$ and from the 
fact that we have $3$ heavy quarks contributing to LO almost with the same 
rate (no enhancement at LO). 
However, the same is not true for the leptons $\Pl'{-}\PGnl'$. 
There are no two-loop diagrams with leptons in $\Pg\Pg\,$-fusion; they enter 
through the renormalization procedure. 
We observe enhancement in the leptonic sector of SM4, which actually dominates 
the behaviour at small values of $\MH$.
To summarize:
%--
\begin{itemize} 
\item SM3 with a heavy-light quark doublet: (positive) enhancement;
\item SM3 with a heavy-heavy (mass-degenerate) quark doublet: (negative) enhancement;
\item SM4 with a heavy-light and heavy-heavy (mass-degenerate) quark doublets: 
enhancement in heavy-light, screening in heavy-heavy;
\item SM4 with a heavy-heavy $\Pl'{-}\PGnl'$ doublet: enhancement.
\end{itemize}
%--
We have verified that our (complete) results confirm the asymptotic estimates of
\Bref{Djouadi:1994ge,Djouadi:1997rj}.

Exclusion of SM4 at LHC requires the most conservative setup and, sometimes, it 
has been suggested to set limits in a scenario where all fermions in the fourth 
generation are ultra-heavy. This is not reasonable for at least three reasons:
there is the usual unitarity requirement which puts a LO bound of approximately
$500\,\UGeV$ (although some recent literature~\cite{Atwood:2011kc} puts the 
current interesting region in the interval $400{-}600\,\UGeV$);
implications of triviality for the Standard Model~\cite{Lindner:1985uk}, where 
one deduces that in the framework of a two-loop renormalization group analysis 
the heaviest quark mass has to be smaller than $400\,\UGeV$. 
This bound emerges from vacuum stability constraints and turns out to be stronger 
than the $500\,\UGeV$ from unitarity.
Finally, EW NLO corrections to $\Pg\Pg$-fusion show that already at 
the $\PQtpr{-}\PQbpr$ threshold, assumed to be at around $1.2\UTeV$, the
contributions of the NLO corrections to the cross-section are as big as the LO one. 

Following the recommendation of the Higgs XS Working Group we have adopted the 
following scenario (see \Brefs{Kribs:2007nz,Hashimoto:2010at}):
%--
\vspace{-.2cm}
\bqa
\Mbp = \Mlp &=& \Mnp = 600\,\UGeV,
\nl
\Mtp - \Mbp &=& 
50 \lpar 1 + \frac{1}{5}\,\ln\frac{\MH}{M_{\rm ref}} \rpar \UGeV\SPp,
\label{const}
\eqa
%--
where $M_{\rm ref} = 115\UGeV$. The reason for this scenario is that while one 
should pay attention to the theoretical upper bounds also lower bounds from 
direct search at LHC matter, so that our assumption will not become obsolete 
in the near future. 
MonteCarlo studies show that the sensitivity at LHC for heavy quark search is 
about $600\,\UGeV$ with $1\,\Ufb^{-1}$ data~\footnote{R.~Tanaka, private 
communication}.
Note that the constraint of \eqn{const} is a severe one and the quarks 
$\PQtpr{-}\PQbpr$ are almost mass-degenerate meaning that, at low values of 
$\MH$, the quarks of the fourth generation contribute very little to the leading 
behaviour of the EW NLO corrections.
%--
\section{Results\label{sec:Results}}
%--
In this Section we present numerical results for complete NLO EW corrections 
to $\Pg\Pg$-fusion in SM4, obtained using the techniques developed 
in~\cite{Actis:2008ts}.

In order to prove that light-fermion dominance in SM3 below $300\,\UGeV$ 
is a numerical accident due to the fact that the top-quark is not heavy enough,
we have computed $\delta^3_{\myEW}$ for a top-quark of 
$800\,\UGeV$ at $\MH= 100\,$GeV and found top-quark dominance 
($\delta^3_{\myEW} = 4.2\%\;(11\%)$ at $\Mt = 172.5\UGeV\;(800\UGeV)$). 
A similar effect in the top-quark sector is also present in SM4; if, for instance, 
we fix all heavy masses to $600\UGeV$, we find $\delta^4_{\myEW} = 12.1\%\;(29.1\%)$ 
at $\Mt = 172.5\UGeV\;(600\UGeV)$.  

\begin{figure}[ht]
\begin{center}
\includegraphics[bb=0 0 567 384,height=6.cm,width=8.cm]{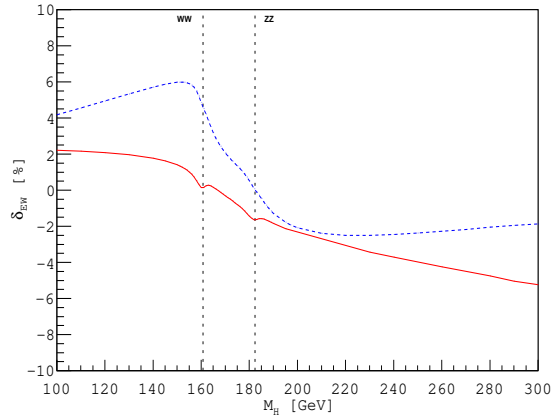}
\end{center}
\vspace{-0.6cm}
\caption[]{\label{Qcomplete} Percentage corrections in SM4
  ($\PQtpr{-}\PQbpr$ quarks only) due to two-loop electroweak
  corrections to $\Pg \Pg \to \PH$ (solid(red) curve). Here we have
  chosen $\Mtp = \Mbp = 400\UGeV$.  For comparison we also show
  the percentage corrections in SM3 (dashed(blue) curve). The vertical dotted
  lines denote the location of the $W{-}W$ and $Z{-}Z$ thresholds.}
\end{figure}

Moving to SM4, the LO $\Pg\Pg$-fusion cross-section is for a light Higgs
boson about 
nine times the one in SM3, \eg see \Bref{Ruan:2011qg}; the contribution from 
pure two-loop EW diagrams should have an impact three times smaller than the 
equivalent one in SM3 in the small $\MH$ region. 
Therefore, if one assumes that also NLO SM4 is dominated by light-fermion 
corrections, \ie that EW corrections are the same for SM3 and SM4, one expects
$\delta^4_{\myEW} \approx \delta^3_{\myEW}/3$ for very heavy fermions. 
According to \Bref{Djouadi:1997rj} this would be true provided that no heavy 
leptons are included.
In \fig{Qcomplete} we show our findings for SM4 in the case where only the quark 
contribution of the fourth generation is included; 
in this case $\delta^4_{\myEW}$ turns to be effectively small for light $\MH$ 
(compared to $\delta^3_{\myEW}$ from \Bref{Actis:2008ug}, dashed(blue) curve), 
but the ratio $\delta^4_{\myEW}/\delta^3_{\myEW}$ is slightly different from the 
expected $1/3$.
This reduction factor applies in fact just to the pure two-loop diagrams, while 
finite renormalization in the top sector remains unchanged moving from 
$\delta^3_{\myEW}$ to $\delta^4_{\myEW}$, giving for the corrections proportional 
to $\Mt^2$ an overall enhancement of a factor $5$ in $\delta^4_{\myEW}$ with 
respect to $\delta^3_{\myEW}$.
%--
\vspace{-.4cm}

\begin{figure}[ht]
\begin{center}
\includegraphics[bb=0 0 567 539,height=6.cm,width=6.cm]{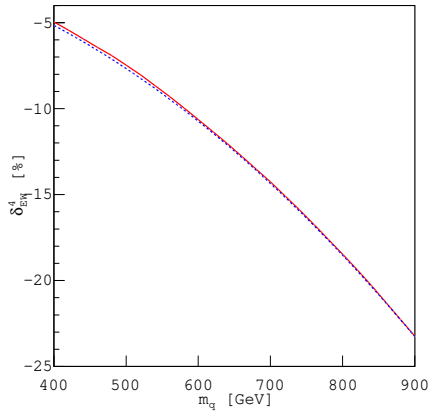}
\end{center}
\vspace{-0.6cm}
\caption[]{\label{quadQ} Comparison of the component of
  $\delta^4_{\myEW}$ due to a mass-degenerate quark isodoublet
  $\PQtpr{-}\PQbpr$ (solid(red) curve) with $-0.7 +
  \frac{4}{3}\,\Delta_{\ssQ}$, where $\Delta_{\ssQ}$ is Eq.(12) of
  \Bref{Djouadi:1994ge}(dashed(blue) curve). The interference of finite
  renormalization $\PQtpr{-}\PQbpr$ quark effects with the one-loop top-quark
amplitude have been excluded.}
\end{figure}

%--
In \fig{quadQ} we have checked our result against the asymptotic limit predicted 
in~\Bref{Djouadi:1994ge}, where the interference between 
finite renormalization effects due to the fourth generation of quarks and the 
one-loop top-quark amplitude was neglected, together with the contribution 
from heavy leptons. 
The solid(red) curve corresponds to our exact result for the EW NLO corrections to 
$\delta^4_{\myEW}$ for fixed $\MH= 100\,\UGeV$ as a function 
of $\MGq = \Mtp = \Mbp$, where the assumptions of~\Bref{Djouadi:1994ge} have 
been applied.
To better appreciate the agreement, the result of~\Bref{Djouadi:1994ge} has been 
shifted (dashed(blue) curve) by a factor $-0.7$, which is our empirical estimate 
of the contributions of the non-enhanced terms, not included in the heavy quark 
expansion of~\Bref{Djouadi:1994ge}.
%--
\vspace{-.4cm}

\begin{figure}[ht]
\begin{center}
\begin{minipage}{6cm}
\includegraphics[bb=0 0 567 539,height=6.cm,width=6.cm]{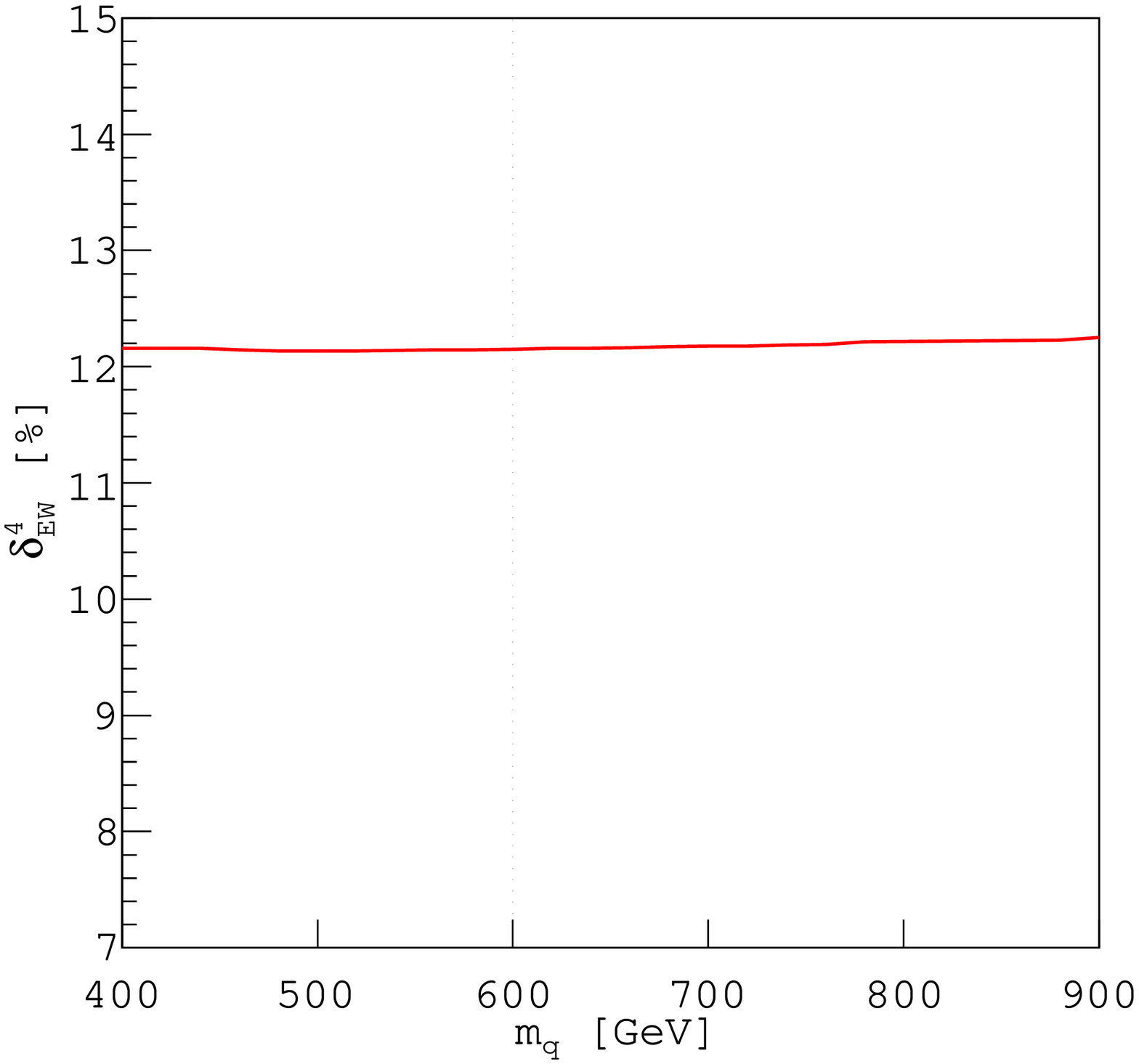}\\[-0.7cm]
\begin{center}
(a)
\end{center}
\end{minipage}
\qquad\qquad
\begin{minipage}{6cm}
\includegraphics[bb=0 0 567 539,height=6.cm,width=6.cm]{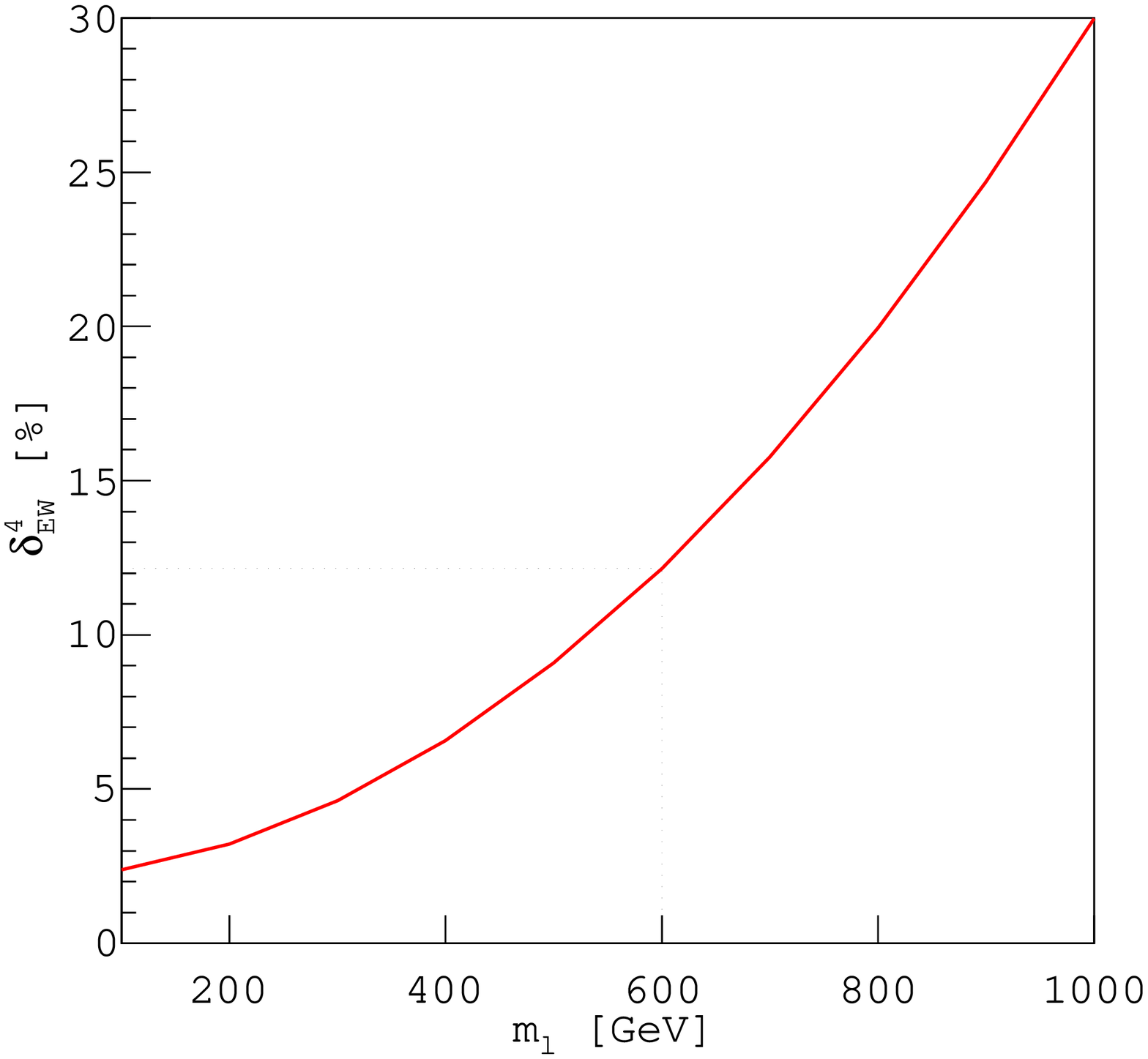}\\[-0.7cm]
\begin{center}
\hspace*{2ex}(b)
\end{center}
\end{minipage}
\end{center}
\vspace{-0.6cm}
\caption[]{\label{UDENdeg}
(Left) Complete $\delta^4_{\myEW}$ due to a degenerate isodoublet 
$\PQtpr{-}\PQbpr$ at $\MH = 100\UGeV$ as a function of 
$\MGq = \Mtp = \Mbp$ for $\Mlp = \Mnp = 600\UGeV$.
(Right) Complete $\delta^4_{\myEW}$ due to a mass-degenerate isodoublet 
$\Pl'{-}\PGnl'$ at $\MH = 100\UGeV$ as a function of 
$\MGl = \Mlp = \Mnp$ for $\Mtp = \Mbp = 600\UGeV$. The vertical line
  denotes the location of the values studied in the following \fig{QLdcomplete}. }
\end{figure}

%--
In \fig{UDENdeg} we consider the dependence of $\delta^4_{\myEW}$ on the masses of 
the fourth generation, at small values of $\MH$.
In the left part is plotted the heavy quark dependence of $\delta^4_{\myEW}$ for a 
mass-degenerate isodoublet $\PQtpr{-}\PQbpr$ at $\MH = 100\UGeV$ as a function of 
$\MGq = \Mtp = \Mbp$ for $\Mlp = \Mnp = 600\UGeV$, showing the expected
screening for mass-degenerate heavy quarks in SM4.
In the right part of \fig{UDENdeg} we plot $\delta^4_{\myEW}$ with a mass-degenerate 
isodoublet $\Pl'{-}\PGnl'$ of leptons at $\MH = 100\UGeV$ as a function of 
$\MGl = \Mlp = \Mnp$ for $\Mtp = \Mbp = 600\UGeV$.
The plot gives complete confirmation of the quadratic enhancement in the masses 
of leptons and neutrinos of SM4.
%--
\vspace{-.4cm}
%--

\begin{figure}[ht]
\begin{center}
\includegraphics[bb=0 0 567 384,height=6.cm,width=9.cm]{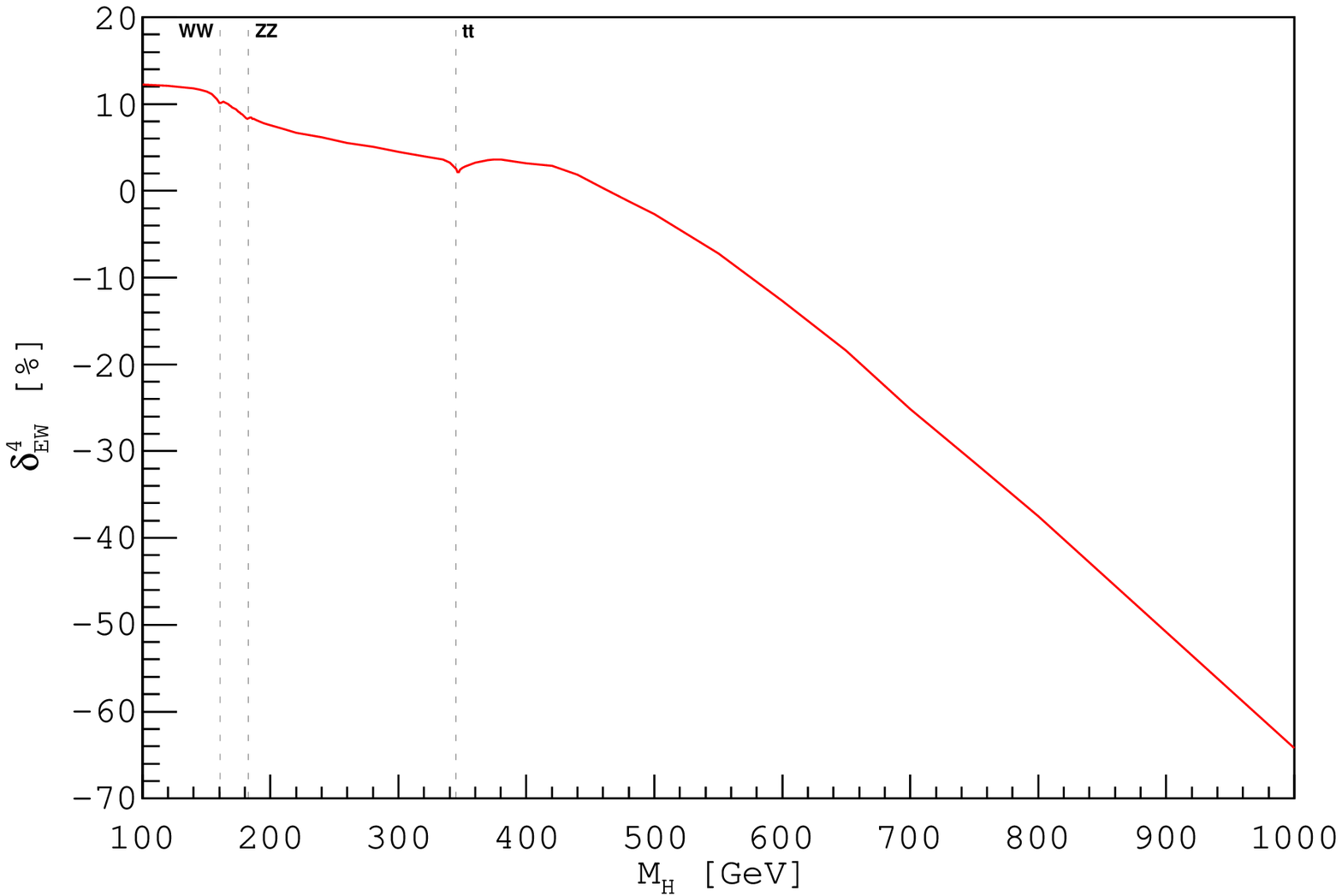}
\end{center}
\vspace{-0.6cm}
\caption[]{\label{QLcomplete} Percentage corrections in SM4
  ($\PQtpr{-}\PQbpr$ and $\Pl'{-}\PGnl'$ doublets) due to two-loop
  electroweak corrections to $\Pg \Pg \to \PH$.  Here we have chosen
  $\Mbp = \Mlp = \Mnp = 600\,\UGeV$ and $\Mtp - \Mbp = 50\,\Big( 1 +
  \frac{1}{5}\,\ln\frac{\MH}{M_{\rm ref}} \Big) \!\UGeV$ with $M_{\rm
    ref}= 115 \UGeV$. The vertical dotted lines denote the location of
  the $W{-}W$, $Z{-}Z$ and $\PAQt{-}\PQt$ thresholds.}
\end{figure}

\begin{figure}[ht]
\begin{center}
\includegraphics[bb=0 0 567 384,height=6.cm,width=9.cm]{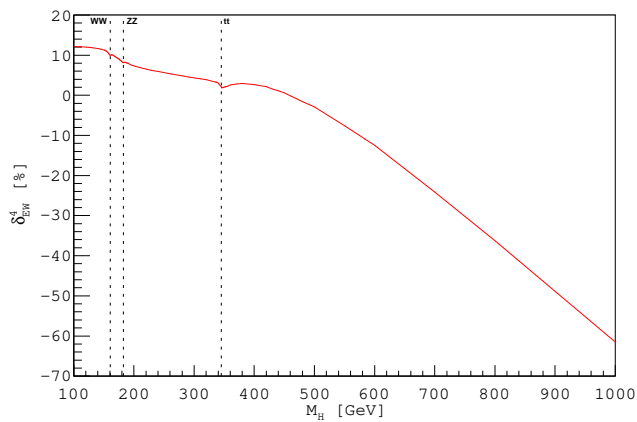}
\end{center}
\vspace{-0.6cm}
\caption[]{\label{QLdcomplete} Percentage corrections in SM4
  ($\PQtpr{-}\PQbpr$ and $\Pl'{-}\PGnl'$ doublets) due to two-loop
  electroweak corrections to $\Pg \Pg \to \PH$.  Here we have chosen
  $\Mtp = \Mbp = \Mlp = \Mnp = 600\,\UGeV$. The vertical dotted lines
  denote the location of the $W{-}W$, $Z{-}Z$ and $\PAQt{-}\PQt$
  thresholds.}
\end{figure}

Our complete result is shown in \fig{QLcomplete} where the $\PQtpr{-}\PQbpr$ and 
the $\Pl'{-}\PGnl'$ doublets are included with the setup of \eqn{const}.
In \fig{QLdcomplete} we consider the alternative scenario with complete 
mass-degenerate fourth generation fermions ($\Mlp = \Mnp = \Mtp = \Mbp = 600\UGeV$).

Electroweak NLO corrections due to the fourth generation are positive and large 
for a light Higgs boson, positive but relatively small around the $\PAQt{-}\PQt$ 
threshold and start to become negative around $450\UGeV$. 
The asymptotic behaviour for $\MH$ well below any heavy $\PAQq{-}\PQq$ threshold 
makes the LO Higgs-gluon-gluon coupling local and allows for a partial(total) 
cancellation of the EW NLO leading corrections due to heavy quark ($\Mt = \Mb$); 
however, the top-quark triangle is the first to become non-local when $\MH$ is 
approaching the $\PAQt{-}\PQt$ threshold, spoiling the asymptotic behaviour.  
Increasing further the value of the Higgs boson mass, the NLO effects tend to 
become huge and negative ($\delta^4_{\myEW} < -100\%$) around the heavy-quark 
threshold.
Above that threshold $\delta^4_{\myEW}$ starts to grow again and becomes positive 
around $\MH= 1750\UGeV$. 
The behaviour at even larger values of $\MH$ shows the usual positive enhancement 
(at $\MH= 3000\UGeV$ we find $\delta^4_{\myEW} > +100\%$), similar to
SM3, entering the non-perturbative regime.
%--
\vspace{-.4cm}
%--
\vspace{-.6cm}

%--
Our result proves that the naive expectation that EW NLO corrections are
dominated by the light fermion contributions fails in SM4. This was
expected for several reasons, including our previous results in SM3
where the partial corrections generated by the top-quark yields
contributions of similar order of magnitude as the one induced by light
fermion loops, smaller though but still in the per cent range as the
whole EW corrections. Comparing \fig{QLdcomplete} with
\fig{UDENdeg}(b), taking into account the $\MGl$-dependence, one can see
that the fourth generation of leptons, which enter the renormalization
procedure, have an important impact on the size of the percentage corrections.
%--
\section{SM4 exclusion\label{sec:Exclusion}}
%--
To put limits on SM4 one also needs to consider the final state; for the low Higgs mass region
the important channels are $\PH \to \PGg\PGg$ and $\PH \to \PZ\PZ^*, \PH \to \PW\PW^*$. In both 
cases it will be difficult to reach exclusion with one channel and it is important to have 
also here the NLO EW corrections in SM4 under control.
Complete NLO EW calculations are not yet available in SM4 for the decay processes and we
can only argue in terms of the leading behaviour of the corrections as given in
\Bref{Djouadi:1997rj}.
All channels show enhancement due to heavy fermions of the fourth generation, however this
leading term is universal for $\PW\PW$ and $\PZ\PZ$ (at least for almost degenerate fermions)
so that we expect less impact on the branching ratios although corrections on partial widths 
are expected to be huge ($\approx -60\%(-20\%)$) for mass-degenerate
fermions $\PQtpr{-}\PQbpr(\Pl'{-}\PGnl')$ of $600\UGeV$.
For $\PH \to \PGg\PGg$ the SM4 branching ratio is suppressed by a factor
$8$ with respect to SM3 but one should be aware of the fact that in SM4
the NLO EW corrections can be as large as the LO partial width.
%--
\section{Conclusions\label{sec:Conclusion}}
%--
In this work we have provided the full two-loop electroweak correction
for the gluon-gluon-fusion process in a Standard Model with a fourth
generation of heavy fermions. Due to the expected enhancement of
radiative corrections we have found a substantially different behaviour
with respect to the same corrections in the Standard Model with three
generations only, also in the region of low Higgs boson masses. The
effect on exclusion limits at LHC are also briefly discussed. Finally,
for values of the heavy-fermion masses given in \eqn{const}
gluon-gluon-fusion becomes non-perturbative in SM4 for values of $\MH$
around $1\UTeV$, as signalled by NLO$\,\approx\,$LO.
%--
\Acknowledgments
%--
We gratefully acknowledge several discussions with A.~De~Roeck, S.~Dittmaier, P.~Gambino, 
C.~Mariotti, R.~Tanaka and M.~Spira.
%--

\providecommand{\href}[2]{#2}\begingroup\raggedright\endgroup

%===
\end{document}